\documentclass{PoS}
\usepackage[font=small,tableposition=top]{caption}

\title{Higgs Bosons in the Standard Model and Beyond}

\ShortTitle{Higgs Bosons in the Standard Model and Beyond}

\author{\speaker{Greg Landsberg}\thanks{On behalf of the ATLAS, CDF, CMS, and D\O\ Collaborations}\\
        Brown University, Department of Physics, 182 Hope St, Providence, RI 02912, USA \\
        E-mail: \email{Greg.Landsberg@cern.ch}}

\abstract{In these proceedings I cover the latest results on the production and decay of the recently discovered Higgs boson. While the spin and properties of the new boson, such as its mass and couplings to bosons and fermions, are covered in a separate report, I focus on individual results in the main channels we use to study the properties of the new boson and to search for its possible cousins, with the focus on the latest results from the LHC and the Tevatron collaborations.}

\FullConference{The European Physical Society Conference on High Energy Physics -EPS-HEP2013\\
		18-24 July 2013\\
		Stockholm, Sweden}

\begin{document}


{\it
I'd like to dedicate this report to the six great minds~--- Francois Englert, Robert Brout, Peter Higgs, Gerald Guralnik, Carl Hagen, and Tom Kibble~--- who made a theoretical breakthrough~\cite{Englert,Higgs,Guralnik} half a century ago, which took so long to confirm experimentally, and Ad Memoriam Robert Brout (1928-2011) who passed on just 14 months before the discovery, which he helped to predict.}

\section{Introduction}

The discovery of the Higgs boson on July 4, 2012~\cite{ATLAS-Higgs,CMS-Higgs} by the ATLAS~\cite{ATLAS} and CMS~\cite{CMS} Collaborations is an apex of experimental particle physics achievements over several past decades. This major discovery would not have been possible if not for the astounding work of thousands of physicists and engineers who built and commissioned the LHC machine~\cite{LHC} and the most sophisticated detectors in particle physics to date: the ATLAS and CMS experiments. Furthermore, the discovery would not have happened so fast if not for enormous progress in particle phenomenology of the last decades, which resulted in very precise calculations of both the Higgs boson production and decay properties, and standard model (SM) backgrounds, which had to be controlled very well in order to uncover this elusive signal. The road to the Higgs boson was paved by the generations of particle physics experiments, most recently the Tevatron ones, which developed many of the analysis techniques that led to the discovery. Finally, we owe this discovery to an excellent LHC machine and the ATLAS and CMS detector performance, which resulted in a large amount of data of very high quality: $\sim 95\%$ of the delivered data were recorded, and $\sim 90\%$ of those were certified and used in the discovery papers and in the subsequent studies covered in this report. 

In the process, both the ATLAS and CMS experiments had to learn how to mitigate the large pileup (multiple interactions per beam crossing) that averaged 14 interactions per beam crossing at the time of the discovery and reached 21 with the second half of 2012 data-taking. The resulting data sample corresponds to an integrated luminosity of approximately 20 fb$^{-1}$ per experiment at the proton-proton center-of-mass energy $\sqrt{s} = 8$ TeV, in addition to an integrated luminosity of $\sim 5$ fb$^{-1}$ collected at $\sqrt{s} = 7$ TeV by each experiment. These combined statistics (on which most of the results presented here are based) are referred to as the LHC Run 1 dataset.

\section{Happy Birthday, Mr. Higgs!}

After the fireworks of July 4 and the announcement of the discovery of a new boson last year, both ATLAS and CMS set off on a long journey of measuring various properties of the new particle and determining if this is the long-sought SM Higgs boson. The amount of work commenced over the year since the discovery is quite remarkable. For most of the channels, the full LHC Run 1 statistics have been analyzed, which amount to 2.5 times the discovery sample. Here are the most important findings of the past year:
\begin{itemize}
\item The existence of the new particle has been established beyond any doubts~\cite{ATLAS-Higgs-new,CMS-Higgs-new} (see Fig.~\ref{fig:ATLASsig} and Table~\ref{table:CMSsig}); 
\item It is a $J^{PC} = 0^{++}$ boson responsible for electroweak symmetry breaking, as evident from its relative couplings to $W/Z$ bosons vs. photons~\cite{ATLAS-Higgs-new,CMS-Higgs-new};
\item Its properties are consistent with those of the SM Higgs boson within (still sizable) uncertainties~\cite{ATLAS-Higgs-new,CMS-Higgs-new}; 
\item There is mounting evidence~\cite{CMS-Higgs-new,Tevatron} that it couples to at least third-generation, down-type fermions, with negative couplings to fermions nearly excluded~\cite{CMS-Higgs-new};
\item The ``big five" channels used in the discovery announcement are about to turn into ``big six", thanks to the new promising results in $t\bar tH(\gamma\gamma,\;b\bar b,\;\tau\tau)$~\cite{ATLAS-ttHgg,CMS-ttHgg,CMS-ttHbbtt}.
\end{itemize}
More details on the properties of the new boson can be found in Ref.~\cite{Fabio} in these proceedings.

\noindent
\begin{minipage}{\textwidth}
\medskip
\begin{minipage}[t]{0.47\textwidth}
\centering
    \vspace{-5.4cm}
    \includegraphics[width=1.0\linewidth]{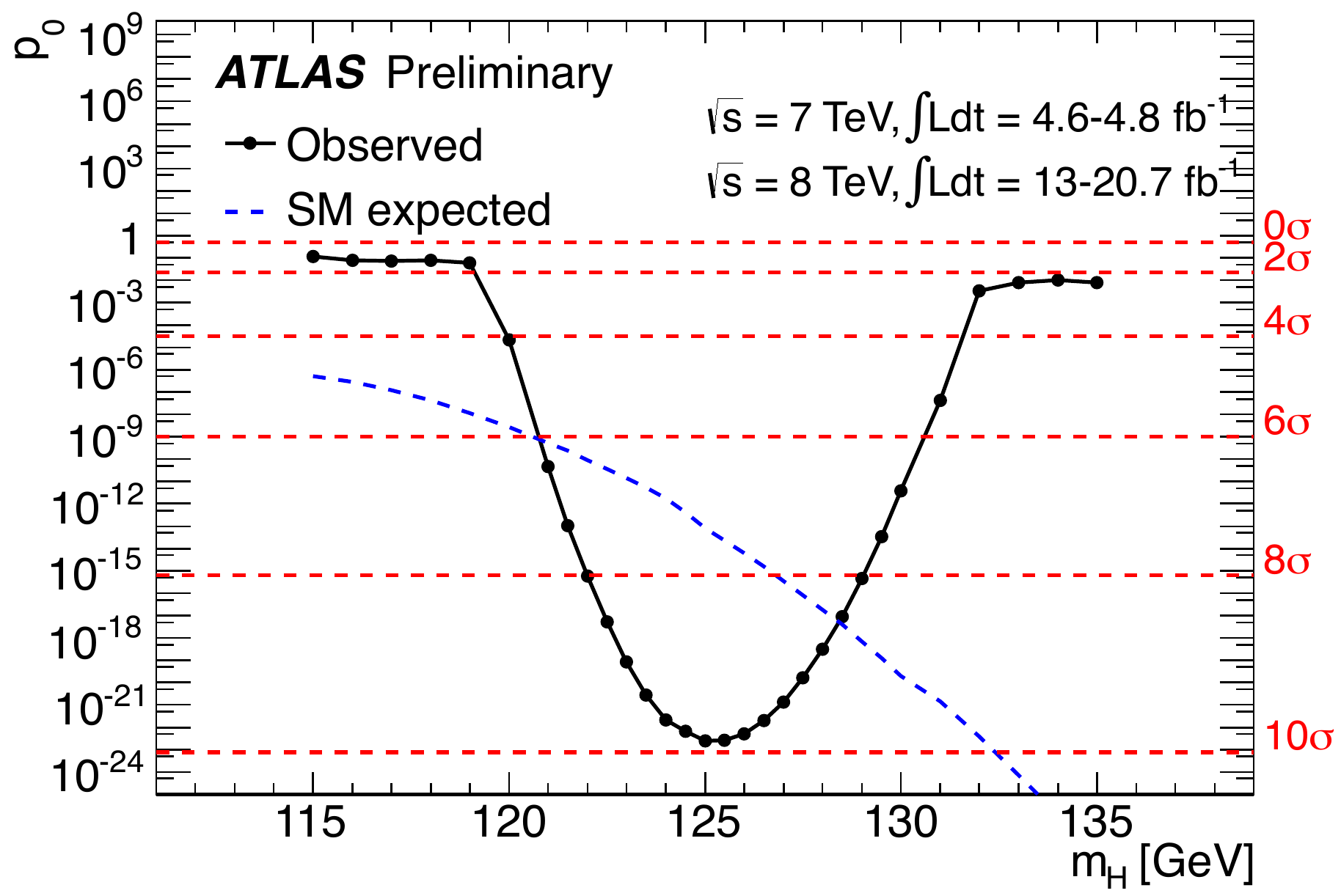}
    \vspace*{-1.0cm}
    \captionof{figure}{Combined Higgs boson observation significance in ATLAS~\protect\cite{ATLAS-Higgs-new}.}
    \label{fig:ATLASsig}
\end{minipage}
\hfill
\begin{minipage}[b]{0.52\textwidth}
\centering
\vspace*{0.0cm}
    \captionof{table}{Expected and observed Higgs boson observation significances in various channels in CMS~\protect\cite{CMS-Higgs-new}.}
    \label{table:CMSsig}
    \vspace*{-0.2cm}
    \begin{tabular}{|l|c|c|}
    \hline
    Channel & Expected, $\sigma$ & Observed, $\sigma$ \\
    \hline
    $ZZ$ & 7.1 & 6.7 \\
    $\gamma\gamma$ & 3.9 & 3.2 \\
    $WW$ & 5.3 & 3.9 \\
    $b\bar b$ & 2.2 & 2.0 \\
    $\tau\tau$ & 2.6 & 2.8 \\
    $b\bar b + \tau\tau$ & 3.4 & 3.4 \\
    \hline
    \end{tabular}
  \vspace{0.6cm}
    \end{minipage}
   \vspace*{-0.4cm}
  \end{minipage}

\section{Higgs Boson Production and Decay}

The following four Higgs boson production mechanisms are currently being probed at the LHC and at the Tevatron: gluon fusion (dominant process with cross section of $\sigma \approx 19.5$ pb for the 125.5 GeV SM Higgs boson produced in $pp$ collisions at $\sqrt{s} = 8$ TeV, and about 5\% of that in $p\bar p$ collisions at $\sqrt{s} = 1.96$~TeV~\cite{Hgg}), vector-boson fusion (VBF), associated production with a vector boson ($VH$), and $t\bar tH$ production.

Once the Higgs boson is produced, it promptly decays via one of many channels, with the dominant branching fractions of 57\%, 22\%, 6.2\%, 2.8\%, and 0.23\% in the $b\bar b$, $WW$, $\tau\tau$, $ZZ$, and $\gamma\gamma$ channels, respectively, for the SM Higgs boson with the mass of 125.5 GeV.

Depending on the production mechanism and the decay channel, the signal-to-background ratio could vary dramatically, so in the following sections we will discuss various combinations of the production and decay modes and their relative importance.

\section{Lucid Higgs}

The most sensitive, high-resolution channel for the Higgs boson measurements is the $gg \to H \to ZZ \to 4\ell$ (with small contribution from other production mechanisms, most notably VBF), where $\ell = e$ or $\mu$. This is a very clean channel and allows for the reconstruction of the Higgs boson mass peak with the resolution of 1--2\%. The high signal-to-background ratio achieved in this channel allows for precise measurements of the Higgs boson mass and spin-parity, as well as its coupling to the $Z$ bosons. In addition, the presence of the Dalitz decays of the $Z$ bosons to four leptons in this channel allows for an {\it in situ\/} calibration of the reconstructed invariant mass scale.

The ATLAS experiment observed the Higgs boson in the $H(ZZ \to 4\ell)$ channel with the significance of $6.6\sigma$ ($4.4\sigma$ expected for the SM Higgs)~\cite{ATLAS-Higgs-new}; while CMS observed it with $6.7\sigma$ significance ($7.2\sigma$ expected)~\cite{CMS-ZZ}. The CMS analysis benefits from higher efficiency for low-$p_T$ leptons, use of angular distributions for improved background rejection, and better muon momentum resolution due to high magnetic field. The signal strength expressed as the ratio of the measured to the SM cross sections ($\mu = \sigma/\sigma_{\rm SM}$) is determined to be $\mu  = 1.43^{+0.40}_{-0.35}$ (ATLAS) and $\mu = 0.91^{+0.30}_{-0.24}$ (CMS), and the mass measured in this channel is $124.3^{+0.6}_{-0.5}\;^{+0.5}_{-0.3}$ GeV (ATLAS) and $125.8 \pm 0.4 \pm 0.2$ GeV (CMS), where the first uncertainty is statistical, and the second one is systematic. This is the most significant (``lucid") channel at the LHC at the moment.

The other sensitive channel is $gg \to H \to \gamma\gamma$, which has very low signal-to-background ratio, but fairly large signal yield. In this channel, ATLAS observed the Higgs boson with the significance of 7.4$\sigma$ ($4.3\sigma$ expected)~\cite{ATLAS-Higgs-new}, while CMS observed it with the significance of $3.2\sigma$ ($4.3\sigma$ expected) in the main, multivariate analysis and $3.9\sigma$ ($3.5\sigma$ expected) in a cross-check analysis based on simpler selections in categories~\cite{CMS-gg}, slightly down from the original significance in this channel at the time of the discovery (the two results are nevertheless consistent within 1.6$\sigma$). The signal strengths measured by ATLAS and CMS are $\mu = 1.55^{+0.33}_{-0.28}$ and $0.78^{+0.28}_{-0.26}$, respectively. The mass measured in this channel is $126.8 \pm 0.2 \pm 0.7$ GeV (ATLAS) and $125.4 \pm 0.8$~GeV (CMS). The CMS mass measurements in the $ZZ$ and $\gamma\gamma$ channels are perfectly consistent with each other, while in the ATLAS case there is a $2.4\sigma$ tension between the two.

In addition, ATLAS has measured~\cite{ATLAS-dsdpt}  several differential cross sections of the Higgs boson production in the $H(\gamma\gamma)$ channel, generally showing good agreement with state-of-the-art theoretical calculations, as shown in Fig.~\ref{fig:dsdpt} (left).

The CMS experiment performed detailed studies of a possibility of a second, nearly degenerate Higgs boson, in the vicinity of 125 GeV, either in the bulk production, or (separately) in the gluon fusion and VBF production~\cite{CMS-Hgg}.

The last ``lucid" channel explored at the LHC is $gg \to H \to WW \to 2\ell + E_T^{\rm miss}$. While this is not a high-resolution channel and, consequently, there is no pronounced mass peak in the spectrum due to missing transverse energy ($E_T^{\rm miss}$) in the event, this is a high-cross-section channel, which serves as an excellent confirmation of the particle observed in the other two ``lucid" final states. The most discriminating variables between the Higgs signal and the background is the dilepton invariant mass and the dilepton transverse mass. The ATLAS analysis~\cite{ATLAS-Higgs-new} uses the latter, while the CMS one~\cite{CMS-WW} explores the 2D distribution in these two variables.

Both collaborations see a clear sign of signal in this channel, consistent with the 125.5 GeV Higgs boson. The observed (expected) significance is $4.0\sigma$ ($3.8\sigma$) in ATLAS~\cite{ATLAS-Higgs-new} and  $4.0\sigma$ ($5.1\sigma$) in CMS~\cite{CMS-WW}. The measured signal strength values are $\mu = 0.99^{+0.31}_{-0.28}$ and $0.76 \pm 0.21$ in ATLAS and CMS, respectively.

In these three lucid channels a clear excess of events has been observed, quantitatively consistent with the expectations from the SM Higgs boson, within $\sim 30\%$ uncertainties.

\section{Visible Higgs}

The other promising decays of the Higgs boson are via the fermonic $b\bar b$ and $\tau\tau$ channels, both of which have sizable branching fractions, some sensitivity to the Higgs boson mass, but are suffering from rather large hadronic backgrounds. Therefore, the search in the $b\bar b$ channel explores an associated $VH$ production, while the $\tau\tau$ search uses sophisticated categorization of the events to enhance the fraction of VBF and associated production categories, but also looking at the gluon fusion process in association with additional jets, which enhances the signal-to-background ratio.

The $VH(b\bar b)$ is also a workhorse channel at the Tevatron. The most recent combination of the CDF and D0 results in this channel~\cite{Tevatron} claims close to $3\sigma$ signal significance with the signal strength consistent with that for the SM Higgs boson, albeit with rather low sensitivity to the Higgs boson mass. In situ calibration with the $VZ$ channel, where the $Z$ boson decays into a pair of $b$-quark jets claims $\sim 3.0\sigma$ $VZ(b\bar b)$ signal, with a budding excess on the high side of the $Z$ mass peak consistent with a SM Higgs boson with the mass measured by the LHC experiments.

\begin{figure}[hbt]
\centering
    \includegraphics[width=0.485\linewidth]{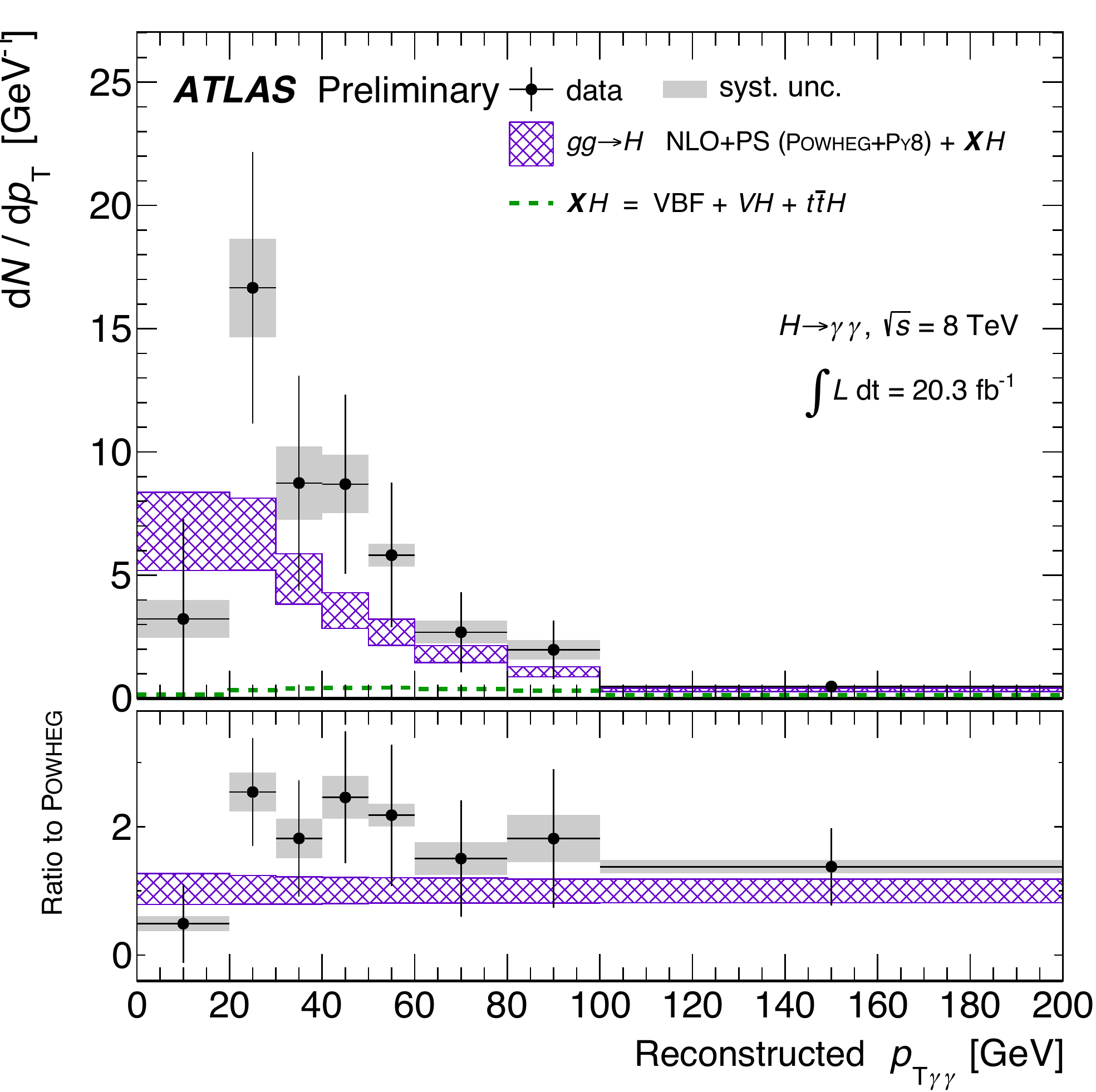}\hfill
    \includegraphics[width=0.514\linewidth]{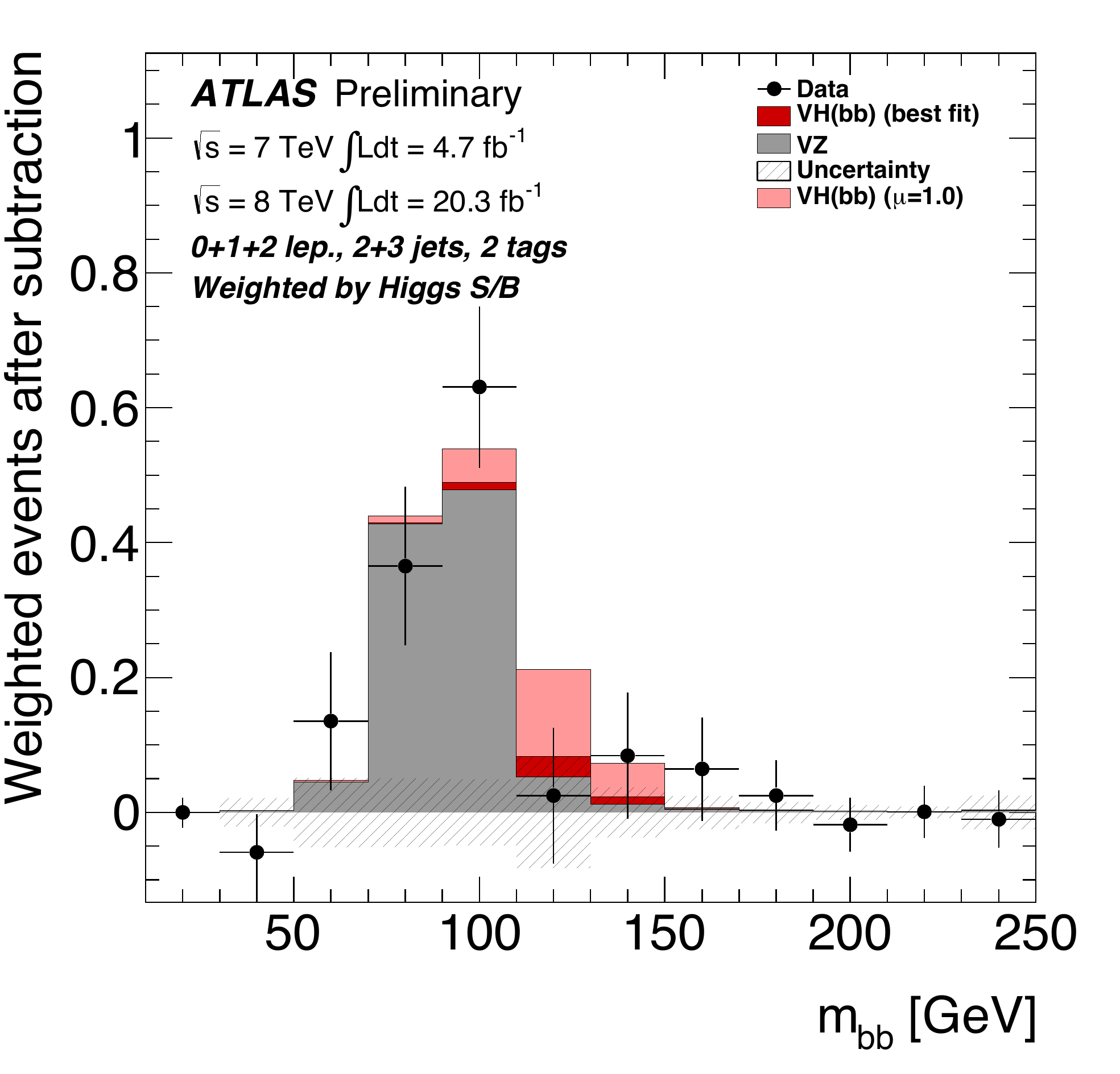}
    \caption{(Left) Higgs boson $p_T$ spectrum in the $gg \to H \to \gamma\gamma$ mode in ATLAS~\protect\cite{ATLAS-dsdpt}.  (Right) Background-subtracted $b\bar b$ mass spectrum from the $VH \to b\bar b$ search in ATLAS~\protect\cite{ATLAS-VHbb}.}\label{fig:dsdpt}
\end{figure}

A recent search from ATLAS~\cite{ATLAS-VHbb} in the $VH(b\bar b)$ channel shows virtually no excess of data over the SM expectations. The measured signal strength in this channel is $\mu = 0.2^{+0.7}_{-0.6}$, with a 95\% confidence level (CL) upper limit on $\mu$ of 1.4 (1.3 expected) for a 125 GeV Higgs boson mass. The $VZ(b\bar b)$ signal is clearly seen in ATLAS with a $4.8\sigma$ significance, as shown in Fig.~\ref{fig:dsdpt} (right).

The analogous CMS analysis~\cite{CMS-VHbb} has achieved a fairly good $b\bar b$ mass resolution of $\sim 10\%$ using the regression techniques. It sees a clear excess over the SM background expectations with the observed and expected significances of $2.1\sigma$. The measured signal strength in this channel is $\mu = 1.00 \pm 0.49$, and the $VZ(b\bar b)$ process has been observed with a $7.5\sigma$ significance.

\begin{figure}[hbt]
\centering
    \includegraphics[width=0.49\linewidth]{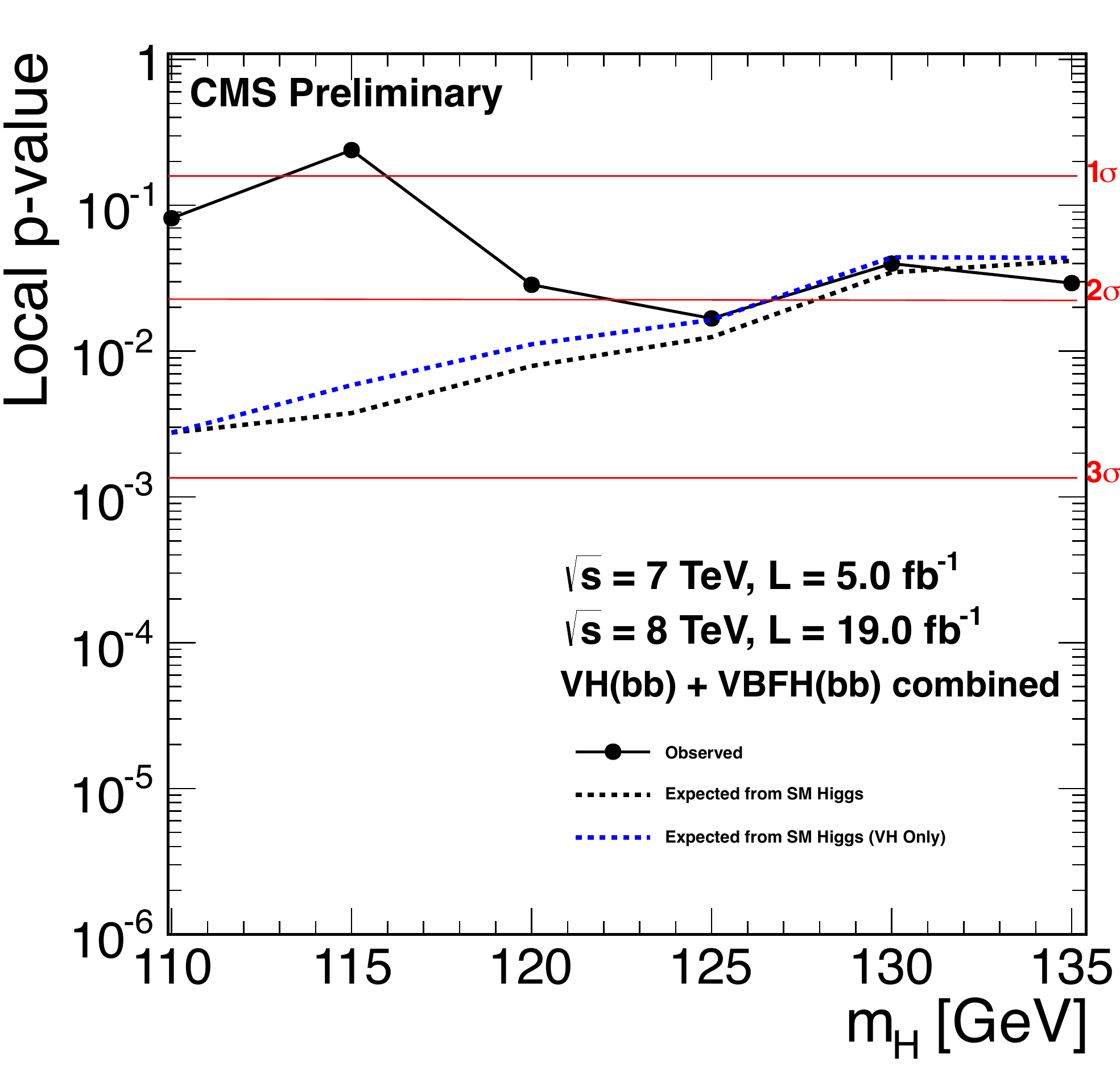}\hfill
    \includegraphics[width=0.49\linewidth]{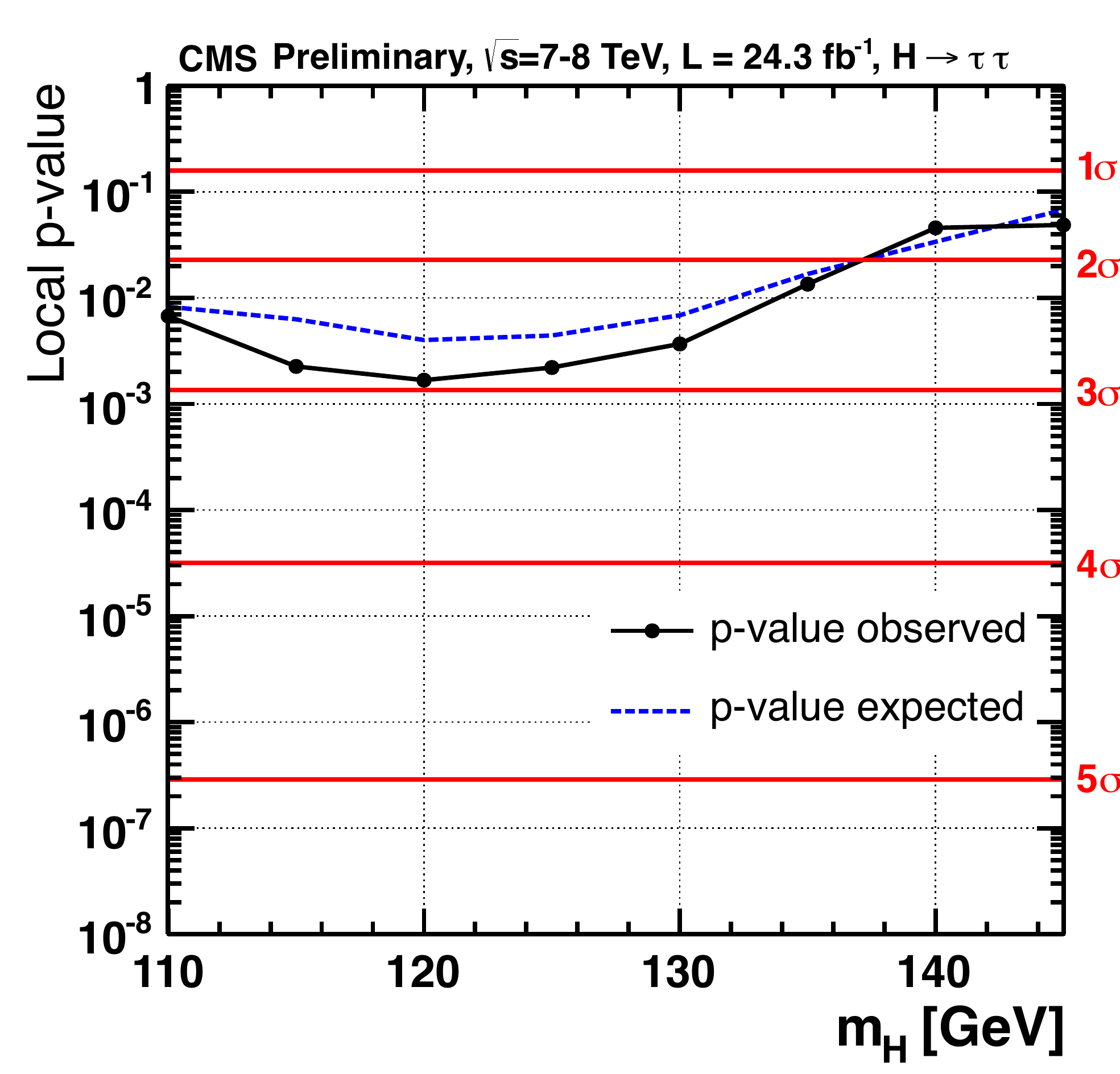}
    \caption{(Left) The expected (dashed black line) and observed (solid line) significance of the combined $VH(b\bar b)$ and VBF $H \to b\bar b$ signal in CMS~\protect\cite{CMS-VHbb,CMS-VBFHbb}. (Right) The expected (dashed) and observed (solid) significance of the $H \to \tau\tau$ signal in CMS~\protect\cite{CMS-Htt}.}
\label{fig:Hbb}
\end{figure}

An interesting new result is the first search for the $H(b\bar b)$ in the VBF production conducted by CMS~\cite{CMS-VBFHbb}. Just as the associated production analysis, the VBF analysis is based on an extensive use of multivariate  techniques to uncover the signal over a large background and reached the sensitivity of $\sim 3$ times the SM cross section for a 125.5 GeV Higgs boson. The combination of this new analysis with the CMS $VH(b\bar b)$ result~\cite{CMS-VHbb} improves the expected sensitivity by about 10\% (while virtually not changing the observed one), and is shown in Fig.~\ref{fig:Hbb} (left).

The ATLAS $H(\tau\tau)$ analysis~\cite{ATLAS-Htt} has not been updated recently and is still based on an integrated luminosity of 4.6 fb$^{-1}$ at 7 TeV and 13.0 fb$^{-1}$ at 8 TeV. It yields the signal strength measurement of $\mu = 0.7 \pm 0.7$ and the observed (expected) signal significance of $1.1\sigma$ ($1.7\sigma$).

The CMS $H(\tau\tau)$ analysis has been updated~\cite{CMS-Htt} to the full LHC Run 1 statistics and benefits significantly from the particle-flow~\cite{PF} reconstruction of $\tau$ leptons, jets, and $E_T^{\rm miss}$. The CMS observed an evidence for the $H(\tau\tau)$ decay with the significance of $2.9\sigma$ ($2.6\sigma$ expected), see Fig.~\ref{fig:Hbb} (right) and measured the signal strength $\mu = 1.1 \pm 0.4$. A combination of this channel with the $VH(b\bar b)$ analysis yields a strong evidence of the Higgs boson coupling to the down-type, third-generation fermions, with the significance of $3.4\sigma$~\cite{CMS-Higgs-new} (see Table~\ref{table:CMSsig}). This is an evidence of at least partial origin of fermion masses from their coupling to the Higgs field. The $H(\tau\tau)$ channel was also used to measure the mass of the new boson to be $120^{+9}_{-7}$ GeV, in a good agreement with the measurements in the high-resolution bosonic decay channels.

\section{Not-yet-Visible Higgs}

Both the ATLAS and CMS collaborations are exploring the possibility to see the Higgs boson signal in other, less sensitive channels. One of such channels is $pp \to H \to Z\gamma$, with the subsequent $Z$-boson decay into electrons or muons. This process is quite sensitive to contributions of new physics, as the decay proceeds through a loop. It also offers a complementary way of searching for new physics in the Higgs boson decay or for non-SM Higgs boson decays, compared to the $pp \to H \to \gamma\gamma$ channel. The SM branching fraction of the $Z\gamma$ decay is only 0.16\% for a 125.5 GeV Higgs boson, which is further suppressed by the leptonic branching fraction of the $Z$-boson decay. Nevertheless, since the decay can be significantly enhanced, it is important to conduct such a search already now. The ATLAS analysis~\cite{ATLAS-Zg} has set an observed (expected) limit on the signal strength $\mu < 18.2$ (13.5), while the CMS analysis~\cite{CMS-Zg} achieved both the observed and expected limit of $\mu < 10$ at a 95\% CL. While the sensitivity is still far away from the SM decay rate, this channel will remain important in the years to come.

Another important rare decay channel is $H \to \mu\mu$, which may be the only way to probe flavor-universality of the Higgs boson couplings in the near future. The SM branching fraction of this decay is only 0.02\% for a SM Higgs boson mass of 125.5 GeV, so this a search remains to be a long shot. Nevertheless, ATLAS has pioneered searches for this rare decay~\cite{ATLAS-mm} and set a limit on the signal strength in this channel of $\mu < 9.8$ (8.6 expected) at a 95\% CL. They also projected~\cite{ATLAS-mm-HL} the sensitivity in this channel for the high-luminosity LHC (HL-LHC) era and showed that a $>6\sigma$ observation is possible with 3 ab$^{-1}$ sample expected at the HL-LHC.

A new player for the Higgs boson studies is the $t\bar tH$ channel, which is vigorously being pursued at the LHC. The suggestion to look for the $t\bar tH$ process predates the LHC operations and originally came from the Tevatron phenomenological study~\cite{ttH}, followed by the ``oscillations" on whether this analysis is possible or not at the LHC. It was clear already from the original paper that this analysis is a tour de force and will require good understanding and control of $t\bar t + X$ backgrounds, many of which were not well calculated theoretically. So, it is not surprising that it took the CDF Collaboration at the Tevatron a whole decade since the original publication to carry out this analysis for the first time~\cite{CDF-ttH}. The first LHC search (from CMS) appeared a year later~\cite{CMS-ttH}, with the sensitivity reaching five times the SM cross section. Today we are getting very close to answering the question of the feasibility of seeing this important signal, which is unique in that it gives an access to the tree-level $t\bar tH$ coupling at the LHC.

\begin{figure}[hbt]
\centering
    \vspace*{-0.1cm}
    \includegraphics[width=0.50\linewidth]{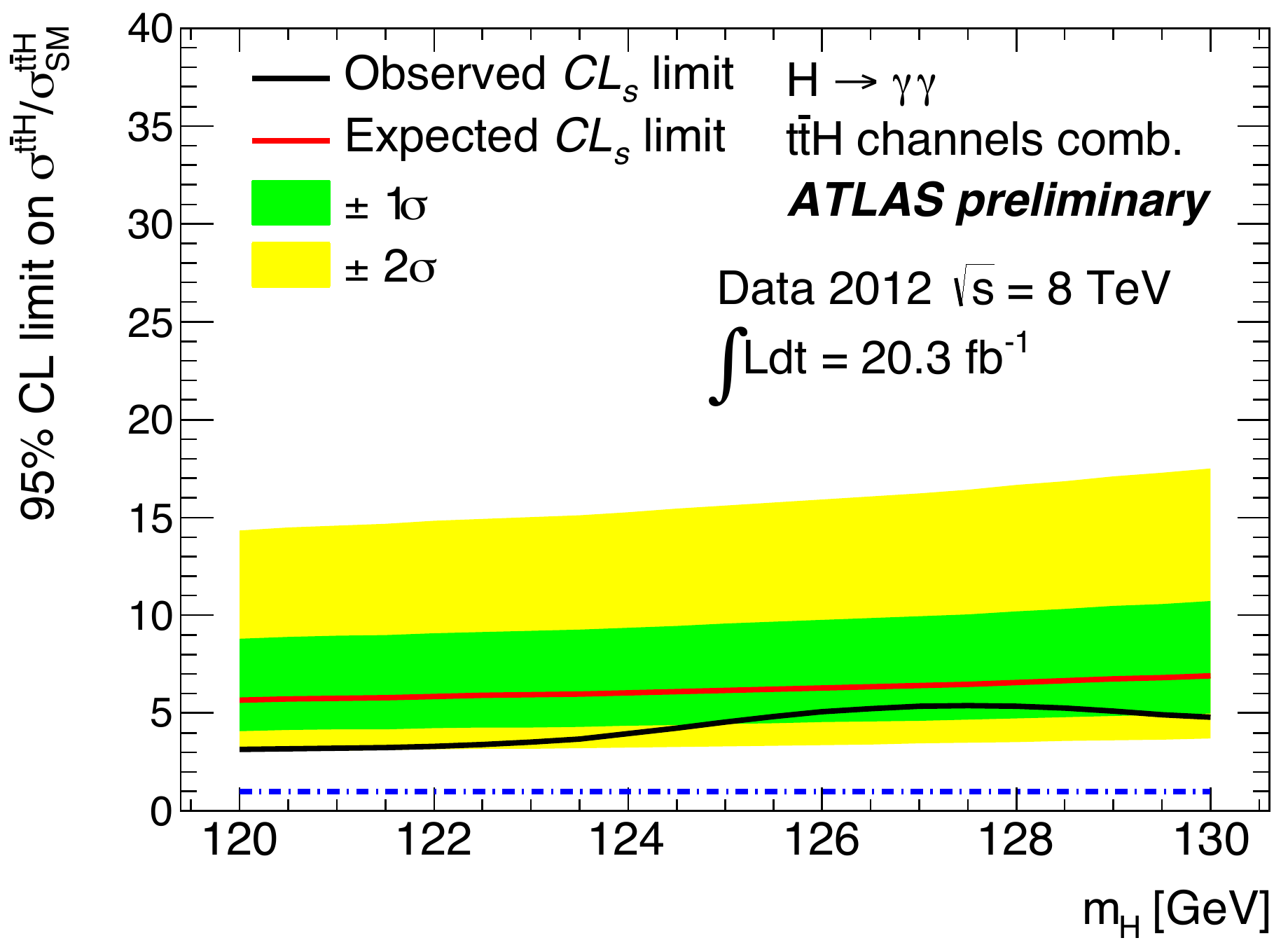}\hfill
    \includegraphics[width=0.50\linewidth]{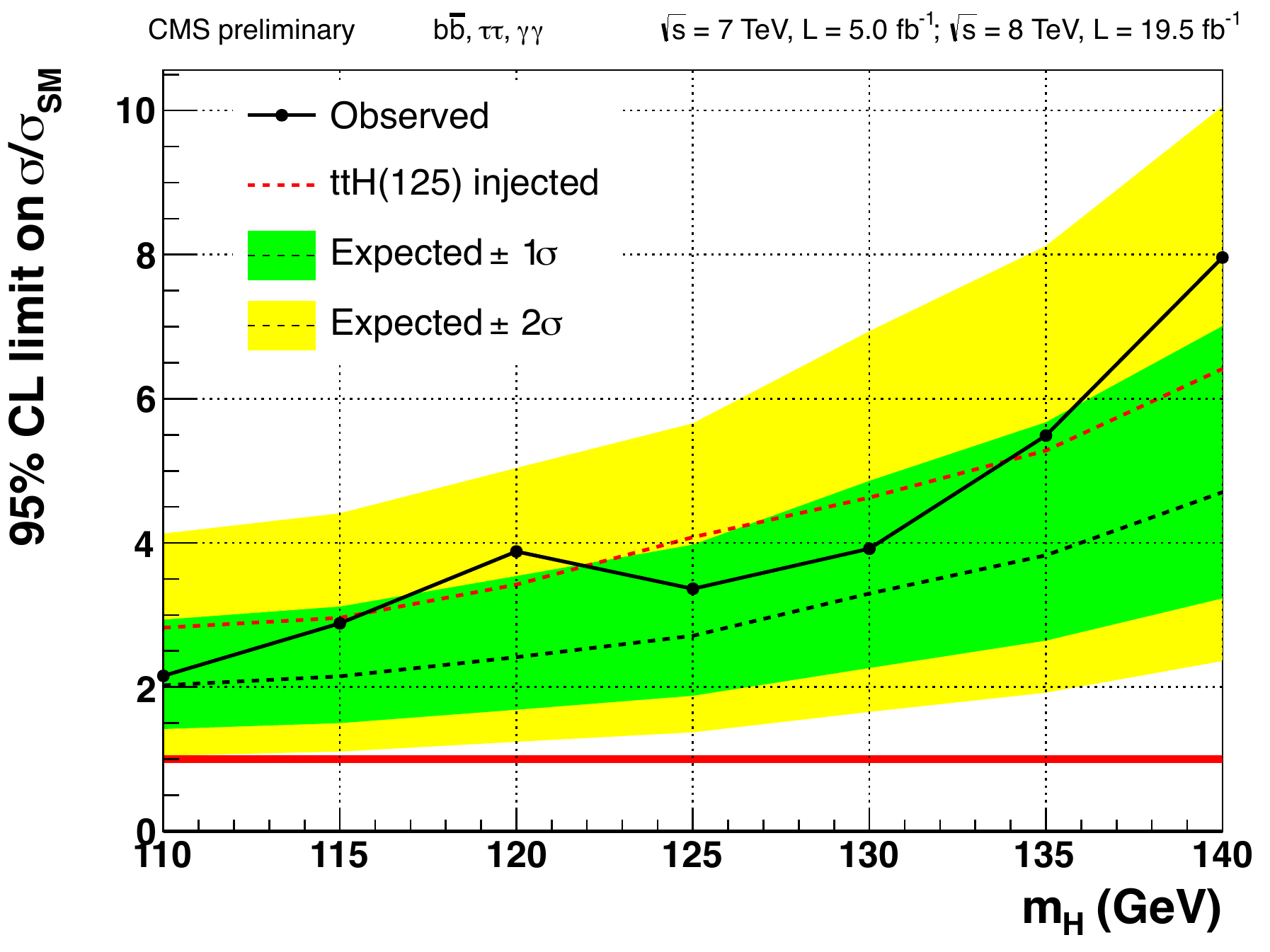}
    \vspace*{-0.7cm}
    \caption{Limits on the $t\bar tH$ production in (left) $H \to \gamma\gamma$ decay mode in ATLAS~\protect\cite{ATLAS-ttHgg} and (right) combination of the $bb$, $\tau\tau$, and $\gamma\gamma$ channels in CMS~\protect\cite{CMS-ttHgg,CMS-ttHbbtt,CMS-ttH}.}
\label{fig:ttH}
\end{figure}

Both ATLAS and CMS experiments released new important $t\bar tH$ results in the $H(\gamma\gamma)$ decay mode. The ATLAS analysis~\cite{ATLAS-ttHgg} set a 95\% CL upper limit on the signal strength of 5.3 (6.4 expected), as shown in Fig.~\ref{fig:ttH} (left), while CMS set a limit of 5.4 (5.3 expected)~\cite{CMS-ttHgg}. The CMS Collaboration has released a new $t\bar tH(b\bar b+\tau\tau)$ analysis~\cite{CMS-ttHbbtt} with full statistics, which sets an upper limit on $\mu$ of 5.2 (4.1 expected) and produced a combination of all three channels with $\mu < 3.4$ (2.7 expected) at a 95\% CL, as shown in Fig.~\ref{fig:ttH} (right).

With an anticipated ATLAS update in the $t\bar tH(b\bar b)$ channel and also ongoing analyses in other channels, one should expect the sensitivity in the $t\bar tH$ channel from the combination of the LHC measurements to reach the SM signal strength, so it is likely that this channel will become the sixth main channel for establishing the Higgs boson existence and could be moved to the previous section of these proceedings soon. I therefore expect it to become a real workhorse channel in the Run 2 of the LHC.

\section{Invisible Higgs}

Another interesting Higgs boson decay channel to explore is an invisible decay, which is vanishingly small in the SM. The interest to this channel is driven by the fact that in many SM extensions the Higgs boson serves as a portal to dark matter (DM) sector, i.e., it is expected to couple to DM particles ($\chi$) and if they are light enough, it could decay invisibly via $H \to \chi\chi$.

Both ATLAS and CMS produced the first direct limits~\cite{invisible} on invisible Higgs boson decay by looking for associated $ZH$ production with the $Z$ boson decaying into a pair of leptons and Higgs boson decaying invisibly, resulting in a dilepton+$E_T^{\rm miss}$ signature. The ATLAS search based on the data corresponding to an integrated luminosity of 4.6 fb$^{-1}$ at 7 TeV and 13.0 fb$^{-1}$ at 8 TeV set a 95\% CL upper limit on the $H \to \chi\chi$ branching fraction of 65\% (84\% expected), while the new CMS analysis based on full LHC Run 1 statistics set a limit of 75\% (91\% expected).

Searches for invisible Higgs boson decays are being pursued in other channels, as well as via indirect constraints coming from the global coupling fits (see Ref.~\cite{Fabio}).

\section{Invincible Higgs}

While the first (and possibly the only) Higgs boson has been found, many extensions of the SM predict more than one Higgs doublet. The best studied class of such models are the two-Higgs-doublet models (2HDM), which have been suggested 40 years ago~\cite{2HDM}, first as an attempt to find a new source of CP violation, and later as a way to separate couplings to up- and down-type fermions. These models have been truly invincible, and their most successful realization~--- supersymmetry (SUSY)~--- continues to elude direct attempts to find it at the LHC, but nevertheless is still a very viable SM extension.

The 2HDM have extended Higgs sector that includes a second CP-even Higgs boson, a CP-odd Higgs boson, and a pair of charged Higgs bosons. An important parameter in 2HDM is $\tan\beta$, which is defined as the ratio of the vacuum expectation values of the two Higgs doublets, as well as $\alpha$, the mixing angle between the two CP-even Higgs bosons, which determines their couplings to fermions. Possible existence of these additional Higgs bosons makes it very important to search for exotic decay channels of the Higgs boson and additional states above or below 125 GeV.

\begin{figure}[hbt]
\centering
    \vspace*{-0.3cm}
    \includegraphics[width=0.43\linewidth]{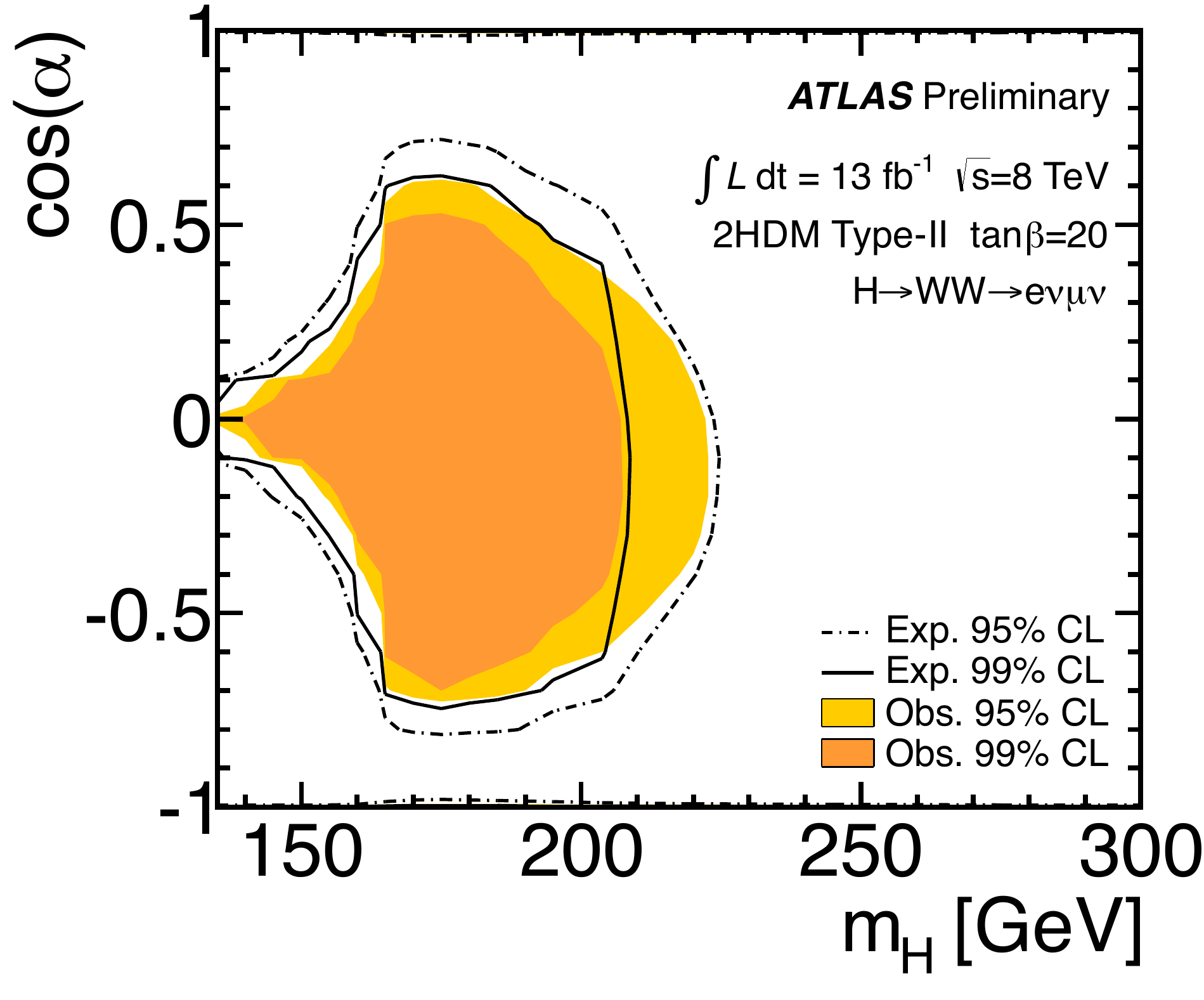}\hfill
    \includegraphics[width=0.57\linewidth]{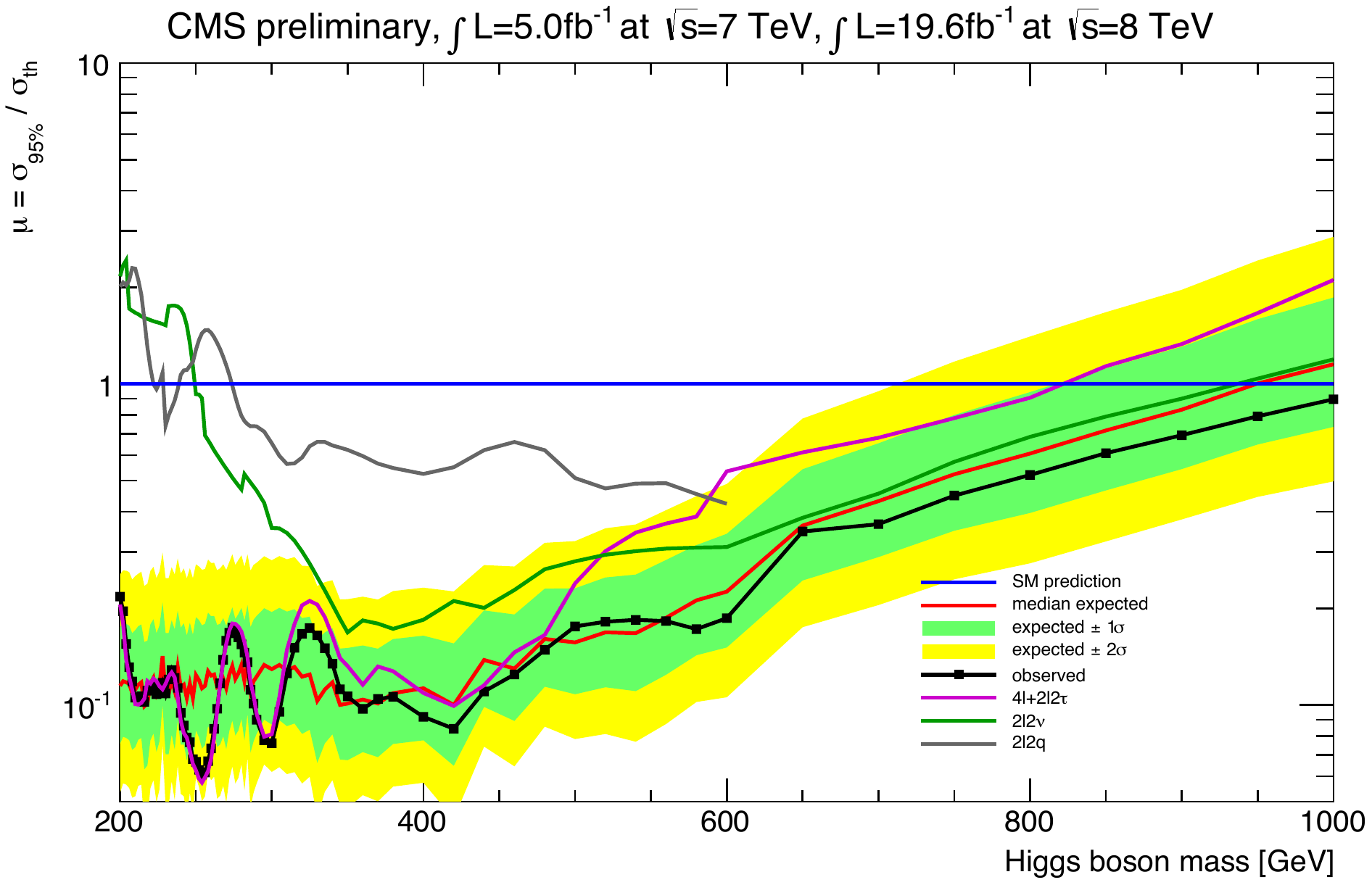}
    \vspace*{-0.7cm}
    \caption{(Left) Limits on Type II 2HDM heavy Higgs boson ($\tan\beta = 20$) from ATLAS~\protect\cite{ATLAS-dsdpt}. (Right) The combined limit on additional Higgs bosons from $H \to ZZ$ channels in CMS~\protect\cite{CMS-ZZ,CMS-ZZ1}.}
\label{fig:heavyH}
\end{figure}

A new ATLAS search~\cite{ATLAS-2HDM} for a heavy CP-even Higgs boson in the $WW$ decay mode considers separately the gluon fusion and VBF production mechanisms as well as Type I (one of the two CP-even Higgs bosons couples to fermions, while the other does not) and Type II (one of the two CP-even Higgs bosons couples to up-type, while the other~--- to down-type fermions) 2HDM. Limits on the second Higgs boson are set in the range of 140-200 GeV, depending on its couplings to fermions and $\tan\beta$ value and are shown in Fig.~\ref{fig:heavyH} (left).

Both ATLAS and CMS searched~\cite{MSSM} for the minimal supersymmetric SM (MSSM) CP-odd Higgs boson in the $\tau\tau$ decay channel. The LHCb collaboration has recently joined the quest using $\tau$ leptons in the forward region~\cite{LHCb}, although it hasn't yet reached the sensitivity of the ATLAS and CMS searches, which rule out $\tan\beta$ above $\sim 5$ for the CP-odd Higgs boson mass between 100 and 200 GeV and $\sim 50$ for the mass of 800 GeV.

Also, stringent limits on a charged Higgs boson from top-quark decays in the $\tau\bar\nu$ (ATLAS and CMS) and $s\bar c$ (ATLAS) channels have been set; as well as limits on Higgs bosons in MSSM extensions that decay in pair of light CP-odd scalars, each of which decays either in a pair of muons (CMS, D\O) or a pair of photons (ATLAS). This program complements earlier searches for light CP-odd scalars in radiative $\Upsilon$ decays at B-factories.

The ATLAS and CMS experiments continue searches for heavy CP-even Higgs bosons with arbitrary couplings in the $\gamma\gamma$, $WW$, $ZZ$, and $t\bar t$ channels. For instance, the recent combination of CMS results in various $H \to ZZ$ decay channels~\cite{CMS-ZZ,CMS-ZZ1} excluded additional Higgs bosons with SM-like couplings to the $Z$ bosons up to the unitarity limit of 1 TeV, as shown in Fig.~\ref{fig:heavyH} (right). For heavy Higgs bosons, some of these searches explicitly explore boosted topology, which results in unresolved decay products of the vector bosons or top quarks~\cite{Higgs-boosted}.

\section{Conclusions}

Higgs physics remains the apex of the LHC physics program. There has been an impressive progress since the discovery of a Higgs boson just a year ago:
\begin{itemize}
\item It is now seen beyond any doubts in three bosonic channels;
\item The spin and the mass of a new state have been determined;
\item It looks more and more like the SM Higgs boson;
\item Coupling to the top quarks has been established indirectly via gluon fusion production mechanism;
\item Couplings to the down-type, third-generation fermions are established at $>3\sigma$ level;
\item No evidence for non-SM Higgs boson decays or additional Higgs bosons at higher or lower mass has been found so far.
\end{itemize}

\begin{figure}[hbt]
\centering
    \includegraphics[width=0.4\linewidth]{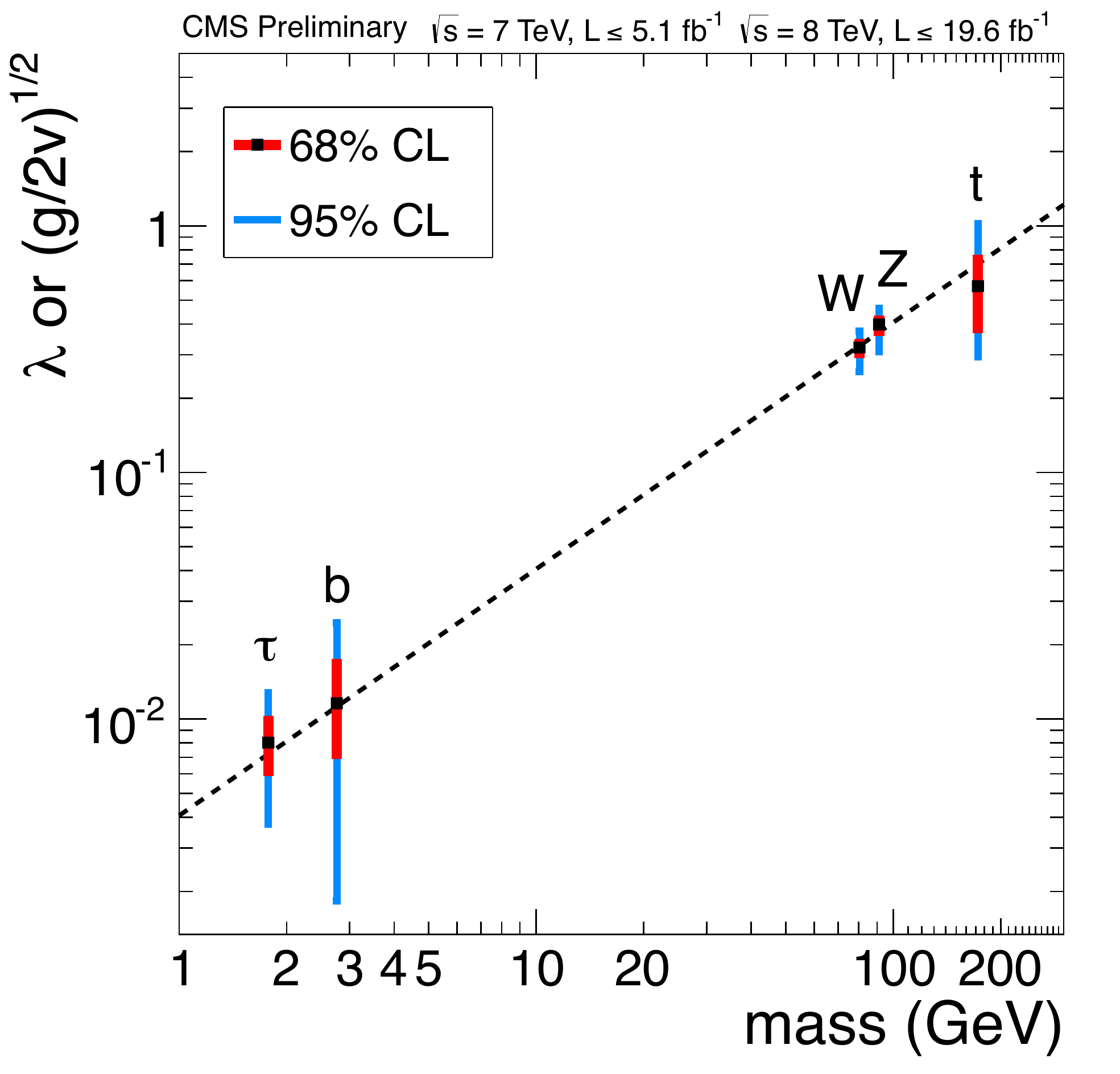}
    \vspace*{-0.4cm}
    \caption{Measurements of the Higgs boson couplings to various SM particles by CMS~\protect\cite{CMS-Higgs-new}.}
\label{fig:Regge}
\end{figure}

Clearly there are many more Higgs physics topics left and many new directions of studies being explored, with an exciting LHC program that will last some two decades, just as the discovery of the top quark in 1995 has opened two decades of beautiful top physics. The main goal for the years to come is to shrink the error bars on the ``Regge plot" (see Fig.~\ref{fig:Regge}) to the dot size and to fill it in with the new entries for muons, Higgs self-coupling, and possibly charm quarks.

\section*{Acknowledgements}

I'd like to thank the organizers for a kind invitation and warm hospitality! I'm also indebted to my many colleagues in ATLAS, CDF, CMS, and D\O\  experiments for producing beautiful results covered in this report and for many helpful discussions. This work has been partially supported by the DOE Grant \#DE-SC0010010.

\end{document}